\documentclass[11pt]{article}

\usepackage{amsmath}

\usepackage[cp1251]{inputenc}

\newcommand{\N}{N\raise.7ex\hbox{\underline{$\circ $}}$\;$}

\begin{document}

\begin{center}
{\bf  E.M. Ovsiyuk\footnote{e.ovsiyuk@mail.ru}, O.V. Veko \\[10mm]
SPIN 1/2 PARTICLE  IN THE FIELD OF THE DIRAC STRING
ON THE BACKGROUND OF ANTI DE SITTER SPACE--TIME} 

{\small  Mozyr State Pedagogical University named after I.P. Shamyakin}

\end{center}

\date{}

\begin{abstract}
The Dirac monopole string is specified for anti de Sitter cosmolo\-gical model.
Dirac equation  for spin 1/2 particle in presence of this monopole has been examined on the background
of anti de Sitter space-time in static coordinates. Instead of spinor monopole har\-monics, the technique of Wigner
$D$-functions is used. After separa\-tion of the variables radial equations have been solved
exactly in terms of hypergeometric functions.
The complete  set of  spinor wave solutions  $\Psi_{\epsilon, j,m, \lambda}(t,r, \theta, \phi)$  has
 been constructed, the most attention is given to treating  the states of minimal values for total
 moment quantum number $j_{min}$. At all values of $j$,  the energy spectrum is discrete.
 

\end{abstract}

\vspace{5mm}

\section{ Introduction}

De Sitter and anti de Sitter geometrical models are given  steady attention in the context
of developing quantum theory in  a curved space-time  -- for instance, see in [1].
In particular, the problem of description  of the particles with different spins  on these curved backgrounds
 has a long history
-- see  [2--34]. Here we will be interested mostly in treating the Dirac equation in  de Sitter model.
In the present paper,  the influence of the  Dirac monopole string  on the spin 1/2 particle
in the anti de Sitter cosmological model
is investigated.
  Instead of spinor monopole harmonics, the technique of Wigner
$D$-functions is used. After separation of the variables radial equation have been solved
exactly in terms of hypergeometric functions.
The complete  set of  spinor wave solutions  $\Psi_{\epsilon, j,m, \lambda}(t,r, \theta, \phi)$  has
 been constructed. Special attention is given to treating  the states of minimal values for total
 moment quantum number $j_{min}$, these states turn to be much more complicated than in the flat Minkowski space.
At all values of $j$,  the energy spectrum is discrete.

\section{ Dirac particle in the anti de Sitter space}

The Dirac equation (the notation according  to \cite{Book-1} is
used)
\begin{eqnarray}
\left [ i \gamma^{c} \; ( e_{(c)}^{\alpha} \partial_{\alpha} + {1
\over 2} \sigma^{ab} \gamma_{abc}) - M \right  ]  \Psi = 0
\label{Dirac}
\end{eqnarray}

\noindent in static coordinates and tetrad of the anti de Sitter space-time
\begin{eqnarray}
 dS^{2} = \Phi \;  dt^{2} - {dr^{2} \over \Phi } -
r^{2} (d\theta^{2} +  \sin ^{2}\theta d\phi ^{2}) \; ,  \; \; \Phi
=  1 + r^{2}  \; , \nonumber
\\
e^{\alpha}_{(0)}=({1 \over  \sqrt{\Phi} }, 0, 0, 0) \; , \qquad
e^{\alpha}_{(3)}=(0, \sqrt{\Phi}, 0, 0) \; , \nonumber
\\
e^{\alpha}_{(1)}=(0, 0, \frac {1}{r}, 0) \; , \qquad
e^{\alpha}_{(2)}=(1, 0, 0, \frac{1}{ r  \sin \theta})  \; ,
\nonumber
\\
\gamma_{030} ={ \Phi ' \over 2\sqrt{ \Phi }} \; , \; \gamma_{311}
={ \sqrt{ \Phi } \over r} \; , \; \gamma_{322} ={ \sqrt{ \Phi }
\over r} \; , \; \gamma_{122} ={ \cos \theta \over r \sin \theta}
\; ,
\end{eqnarray}

\noindent
takes the form
\begin{eqnarray}
\left [  i {\gamma ^{0}  \over \sqrt{\Phi}} \partial _{t}  + i
\sqrt{\Phi } \left (     \gamma ^{3}
\partial _{r}  + { \gamma ^{1} \sigma^{31}  +
\gamma  ^{2} \sigma^{32} \over r }   +\right. \right.
\nonumber
\\
\left. \left.   { \Phi' \over 2 \Phi  } \gamma ^{0}     \sigma^{03} \right  )  +
 { 1 \over r} \Sigma _{\theta,\phi }  -  M
 \right  ]    \Psi (x)  =  0 \; ,
 \label{10.1a}
 \end{eqnarray}

 \noindent where
 \begin{eqnarray}
 \Sigma _{\theta ,\phi } =  \; i\; \gamma ^{1}
\partial _{\theta } + \gamma ^{2} {i\partial +i\sigma^{12}\cos
\theta  \over \sin \theta  } \; . \nonumber
\end{eqnarray}

\noindent  Eq. (\ref{10.1a}) reads
\begin{eqnarray}
\left [ i\; {\gamma ^{0}  \over \sqrt{\Phi} }   \partial _{t} \; + i
\sqrt{\Phi } \gamma^{3}\; (  \;
\partial _{r}  +   { 1 \over r }   +
  { \Phi' \over 4 \Phi  }  \;  )
+   { 1 \over r} \;\Sigma _{\theta,\phi } -  M
  \right ]   \Psi (x)  =  0  \; .
\label{10.1b}
\end{eqnarray}

\noindent From (\ref{10.1b}), with the
substitution
$
 \Psi  (x)  = r ^{-1} \Phi ^{-1/4} \; \psi (x)$,
 we get
\begin{eqnarray}
\left ( i\; {\gamma ^{0}  \over \sqrt{\Phi} }   \partial _{t} \; + i
\sqrt{\Phi } \gamma^{3}\;   \;
\partial _{r}  +   { 1 \over r} \;\Sigma _{\theta,\phi }  -  M \;
  \right )   \psi (x)  =  0  \; .
\label{10.1c}
\end{eqnarray}

\noindent Below the spinor basis will be used
\begin{eqnarray}
\gamma^{0} = \left | \begin{array}{cc}
0 & I \\
I & 0
\end{array} \right |, \;  \gamma^{j} =
\left | \begin{array}{cc}
0 & -\sigma_{j} \\
\sigma_{j}  & 0
\end{array} \right |, \;  i \sigma^{12} =
\left | \begin{array}{cc}
\sigma_{3} & 0 \\
0 & \sigma_{3}
\end{array} \right | .
\nonumber
\end{eqnarray}

\section{   Separation of the variables }

Let us start with  the
monopole Abelian potential in the Schwinger's form \cite{Strazhev-Tomilchik} in flat Minkowski space
\begin{eqnarray}
A^{a}(x) = (A^{0}, \; A^{i}) = \left ( 0 \; , \; g\;
 {(\vec{r} \times \vec{n})\;(\vec{r} \; \vec{n}) \over
r \; (r^{2} - (\vec{r} \; \vec{n})^{2}) } \right )  .
\label{3.1a}
\end{eqnarray}

\noindent Specifying $\vec{n} = (0, 0 , 1 )$ and translating  the $A_{\alpha }$  to  the~spherical
coordinates,  we get
\begin{eqnarray}
A_{0} = 0 \; , \;\; A_{r} = 0 \; , \;\; A_{\theta } = 0\; , \qquad
A_{\phi } = g\; \cos \theta  \; .
\label{3.1b}
\end{eqnarray}

\noindent It is easily verified that this potential$A_{\phi}$ obeys
Maxwell equations in anti de Sitter space
\begin{eqnarray}
{1 \over \sqrt{-g}} {\partial \over \partial x^{\alpha}} \sqrt{-g}  F^{\alpha \beta} = 0\; ,
\sqrt{ -g } = r^{2} \sin \theta\;,
\nonumber
\\
F_{\phi \theta} = g \sin \theta \; ,
\qquad
{1 \over r^{2} \sin \theta }  {\partial \over \partial \theta}  r^{2} \sin \theta\;
{1 \over r^{2}}  {1 \over  r^{2}\sin^{2} \theta} \;  g \sin \theta =0 \; .
\label{3.1c}
\end{eqnarray}

Correspondingly,  the~Dirac  equation  in  this   electromagnetic
field  takes the form
\begin{eqnarray}
\left [ i\; {\gamma ^{0}  \over \sqrt{\Phi} }   \partial _{t} \; + i
\sqrt{\Phi } \gamma^{3}\;   \;
\partial _{r}  +   { 1 \over r} \;\Sigma^{k} _{\theta,\phi }  -  M \right
  ]   \psi (x)  =  0  \; ,
\label{3.2a}
\end{eqnarray}

\noindent where
\begin{eqnarray}
\Sigma ^{k}_{\theta ,\phi }  =  i \gamma ^{1} \partial _{\theta}   +
\gamma ^{2}   { i \partial _{\phi } + (i\sigma ^{12} - k )
 \cos \theta \over  \sin  \theta} \; ,
\label{3.2b}
\end{eqnarray}

\noindent and $k \equiv  eg/hc$.
As readily verified, the~wave  operator
  in  (\ref{3.2a}) commutes with the~following three ones
\begin{eqnarray}
J^{k}_{1} =  l_{1} + {(i\sigma ^{12} - k)
 \cos \phi  \over \sin \theta } \; ,
 \nonumber
 \\
J^{k}_{2} = \ l_{2} + {(i\sigma ^{12} - k)
\sin \phi  \over \sin \theta } \; , \;\;\;
\qquad  J^{k}_{3} = l_{3}\; ,
\label{3.3a}
\end{eqnarray}

\noindent which     obey   the~$su(2)$   Lie   algebra.   Clearly,
this    monopole     situation     come  entirely  under
the~Schr\"{o}dinger
\cite{Schrodinger-1938}  and Pauli
\cite{Pauli-1939} approach (detailed treatment of the method was given in \cite{Book-2}).
Correspondingly to diagonalizing  the
 $\vec{J}^{2}_{k}$ and $J^{k}_{3}$,
the~function $\psi$ is to be taken as
($D_{\sigma } \equiv D^{j}_{-m,\sigma }(\phi ,\theta ,0)$ stands for Wigner functions
\cite{Varshalovich-Moskalev-Hersonskiy-1975})
\begin{eqnarray}
\psi^{k}_{\epsilon jm} (t,r,\theta ,\phi ) = e^{-i\epsilon t}
\left |\begin{array}{r}
       f_{1} \; D_{k-1/2}   \\   f_{2} \; D_{k+1/2}   \\
       f_{3} \; D_{k-1/2}   \\   f_{4} \; D_{k+1/2}
\end{array} \right |\; .
\label{3.3b}
\end{eqnarray}

\noindent Further, with the he;p of recursive relations  \cite{Varshalovich-Moskalev-Hersonskiy-1975}
\begin{eqnarray}
\partial_{\theta}   D_{k+1/2} =  a  D_{k-1/2} - b  D_{k+3/2}
\; ,\;
\partial_{\theta}   D_{k-1/2} =  c  D_{k-3/2} - a  D_{k+1/2} \; ,
\nonumber
\\
\sin^{-1} \theta   \;[\; -m -(k+1/2) \cos \theta \;] \; D_{k+1/2} =
(- a  D_{k-1/2} - b  D_{k+3/2} )  \;\; ,
\nonumber
\\
\sin^{-1} \theta  \;[\; -m -(k-1/2)\cos \theta \;] \; D_{k-1/2} =
(- c  D_{k-3/2} - a  D_{k+1/2} )\; ,
\nonumber
\end{eqnarray}
\begin{eqnarray}
b = { \sqrt{(j - k - 1/2)(j + k + 3/2)}  \over 2}\; ,
\nonumber
\\
 c = { \sqrt{(j + k - 1/2)(j - k + 3/2)}  \over 2}\; ,
\nonumber
\\
a = {1 \over 2} \sqrt{(j + 1/2)^{2} - k^{2}}
\nonumber
\end{eqnarray}

\noindent
we find how the $\Sigma ^{k}_{\theta ,\phi }$  acts on $\psi ^{k}_{\epsilon jm} $
\begin{eqnarray}
\Sigma ^{k}_{\theta ,\phi } \; \psi ^{k}_{\epsilon jm} =
 i  \sqrt{(j + 1/2)^{2} - k^{2}} \;\; e^{-i\epsilon t} \;
\left | \begin{array}{r}
  - f_{4} \; D_{k-1/2}  \\  + f_{3} \; D_{k+1/2} \\
  + f_{2} \; D_{k-1/2}  \\  - f_{1} \; D_{k+1/2}
\end{array} \right |;
\label{3.4}
\end{eqnarray}

\noindent
hereafter the factor $\sqrt{(j + 1/2)^{2}- k^{2}}$
 will be denoted by $\nu $.
 For the  $f_{i}(r)$, the radial system derived is
 \begin{eqnarray}
{\epsilon  \over \sqrt{\Phi}} \; f_{3} \; - \; i \; \sqrt{\Phi} {d\over dr} \; f_{3}  \;- \;i\; {\nu \over r}\; f_{4}\;
 - \; M \; f_{1} = 0 \; ,
\nonumber
\\
{\epsilon  \over \sqrt{\Phi}} \; f_{4} \; + \; i\; \sqrt{\Phi}  {d \over dr}\;  f_{4}\; + \;
 i \;{\nu \over r} \; f_{3} \; - \; M \; f_{2} = 0 \; ,
\nonumber
\\
{ \epsilon \over \sqrt{\Phi}} \; f_{1} \; +\; i\; \sqrt{\Phi}{d \over dr} \; f_{1} \;  + \;
i \; {\nu \over r} \; f_{2} \;- \;M\; f_{3} = 0 \; ,
\nonumber
\\
{ \epsilon \over \sqrt{\Phi}} \; f_{2} \; - \;i\; \sqrt{\Phi} {d \over dr}\; f_{2}\; - \;
i \;{\nu \over r}\; f_{1}\; -\; M \;f_{4} = 0 \; .
\label{3.5}
\end{eqnarray}

Else  one  operator  can  be
diagonalized together with $ i \partial _{t} , \vec{J}^{2}_{k},
J^{k}_{3} $:  namely, a~generalized Dirac operator
\begin{eqnarray}
\hat{K} ^{k} \; = - \; i \; \gamma^{0} \; \gamma ^{3} \;
\Sigma ^{k}_{\theta ,\phi }   \; .
\label{3.6a}
\end{eqnarray}

\noindent From the  equation $\hat{K}^{k} \psi _{\epsilon jm}  =
\lambda  \; \psi _{\epsilon jm}$    we  find   two
possible eigenvalues  and  restrictions  on $f_{i}(r)$
\begin{eqnarray}
f_{4} = \delta \;  f_{1} \; , \;\;\; f_{3} = \delta \; f_{2} \;,
\qquad
\lambda  = - \delta \;  \sqrt{(j + 1/2)^{2}- k^{2}} \;  .
\label{3.6b}
\end{eqnarray}

\noindent Correspondingly,  the~system  (\ref{3.5})  reduces to
\begin{eqnarray}
\left ( \sqrt{\Phi} {d \over dr} + {\nu \over r} \right ) f \; + \;
\left ({\epsilon  \over \sqrt{\Phi}} + \delta\;  M  \right )\;  g = 0\; ,
\nonumber
\\
\left ( \sqrt{\Phi} {d \over dr} - {\nu \over r} \right ) g \; - \;
\left  ({\epsilon  \over \sqrt{\Phi}}  - \delta\;  M  \right ) \; f = 0\; ,
\label{3.7}
\end{eqnarray}

\noindent
where
\begin{eqnarray}
f ={f_{1} + f_{2} \over \sqrt{2}}\;, \qquad g =  {f_{1} - f_{2} \over \sqrt{2} i } \; .
\nonumber
\end{eqnarray}

It is known that quantization of  $k = eg/hc$  and $j$ is given by
\begin{eqnarray}
eg /  hc  = \pm 1/2 , \; \pm 1, \; \pm 3/2, \ldots ;
\nonumber
\\
j = \mid k \mid  -1/2, \mid k \mid +1/2, \mid k \mid +3/2,\ldots
\label{3.8}
\end{eqnarray}

The case of minimal  value $j_{min}= \mid k \mid - 1/2$  must  be
treated separately in a  special way.  For example, let
$k = +1/2$, then to the minimal value $j = 0$ there corresponds the~wave
function in terms of only $(t,r)$-dependent quantities
\begin{eqnarray}
\psi ^{(j=0)}_{k = +1/2}(x) =  e^{-i\epsilon t}
\left | \begin{array}{c}
           f_{1}(r)  \\   0  \\  f_{3}(r)  \\  0
\end{array} \right |   \; .
\label{3.9a}
\end{eqnarray}

\noindent At $k = - 1/2$,   we have
\begin{eqnarray}
\psi ^{(j =0)}_{k = -1/2}(x) =
e^{-i\epsilon t}
\left | \begin{array}{c}
   0  \\  f_{2}(r)  \\   0   \\  f_{4}(r)
\end{array} \right |       \;          .
\label{3.9b}
\end{eqnarray}

\noindent Thus, if $k = \pm  1/2$, then to the minimal  values
 $j_{\min }$
there correspond the function substitutions which do not depend at
all on the angular variables $(\theta ,\phi )$; at
 this point there exists some
formal analogy between  these  electron-monopole  states  and
$S$-states (with $l = 0 $) for a~boson field of spin zero:
$\Phi _{l=0} = \Phi (r,t)$. However, it would be unwise to attach too much
significance
to this formal similarity  because that $(\theta ,\phi )$-independence
of $(e-g)$-states  is  not a fact  invariant   under   tetrad   gauge
transformations.

In contrast, the relation below (let $k = +1/2)$
\begin{eqnarray}
\Sigma^{+1/2}_{\theta ,\phi } \; \psi ^{(j=0)}_{k=+1/2} (x)  =
\gamma ^{2}  \cot \theta \; ( i \sigma ^{12} - 1/2 ) \;
\psi ^{(j =0)} _{k=+1/2} \equiv  0
\label{3.10a}
\end{eqnarray}

\noindent is invariant under arbitrary tetrad gauge transformations.
Correspondingly, the~matter equation (\ref{3.2a}) takes on the form
\begin{eqnarray}
\left (   i \; {\gamma ^{0} \over \sqrt{\Phi} } \; {\partial \over \partial t}  +
 i\; \gamma ^{3} \sqrt{\Phi}  \; {\partial \over \partial r}
 -   M  \right )  \psi ^{(j=0)} = 0\; .
\label{3.10b}
\end{eqnarray}

It is readily  verified  that  both (\ref{3.9a})  and (\ref{3.9b}) representations are
directly extended to $(e-g)$-states  with $j = j_{\min }$ at all  the other
$k =\pm 1, \pm 3/2, \ldots $ Indeed,
\begin{eqnarray}
k= +1, +3/2, +2,\ldots, \qquad
\psi ^{k > 0} _{j_{min.}} (x) = e^{-i\epsilon t}
\left | \begin{array}{l}
   f_{1}(r) \; D_{k-1/2}  \\  0  \\  f_{3}(r) \;  D_{k-1/2} \\  0
\end{array} \right | \; ;
\label{4.11a}
\\
k = -1, -3/2,-2,\ldots,  \qquad
\psi  ^{k<0} _{j_{min.}} (x) =  e^{-i\epsilon t}
\left | \begin{array}{l}
    0    \\   f_{2}(r) \; D_{k+1/2}  \\  0  \\ f_{4}(r) \; D_{k+1/2}
\end{array} \right | \; ,
\label{4.11b}
\end{eqnarray}

\noindent and  the relation
$\Sigma _{\theta ,\phi } \psi _{j_{\min }} = 0 $ still  holds.
For instance, let us consider in more detail  the case of positive $k$.
Using the recursive relations
\begin{eqnarray}
\partial _{\theta } D_{k-1/2} =
 { 1 \over 2} \sqrt{ 2k-1}  \; D_{k-3/2}\; ,
 \nonumber
 \\
\sin^{-1} \theta \; [ \;  - m - (k-1/2) \cos \theta \; ] \;   D_{k - 1/2}  =
 - { 1 \over 2} \sqrt{ 2k -1} \; D_{k-3/2} \; ,
\nonumber
\end{eqnarray}

\noindent we get
\begin{eqnarray}
i\gamma ^{1} \; \partial _{\theta}
\left |  \begin{array}{c}
       f_{1}(r) \; D_{k-1/2} \\  0  \\  f_{3}(r) \; D_{k-1/2}  \\  0
\end{array} \right | = {i\over 2} \sqrt{2k-1}
\left | \begin{array}{c}
     0  \\ - f_{3}(r) \; D_{k-3/2}  \\ 0  \\ + f_{1}(r)\; D_{k-3/2}
\end{array} \right | ,
\nonumber
\\
\gamma ^{2}  \; {i\partial _{\phi } + (i\sigma ^{12} - k) \cos \theta \over
\sin \theta} \;
\left | \begin{array}{c}
      f_{1}(r) \; D_{k-1/2} \\  0  \\  f_{3}(r) \; D_{k-1/2} \\ 0
\end{array} \right |  =
{i \over 2} \sqrt{2k-1}
\left | \begin{array}{c}
    0 \\ +f_{3}(r) \; D_{k-3/2}  \\ 0  \\ -f_{1}(r) \; D_{k-3/2}
\end{array} \right | ;
\nonumber
\end{eqnarray}

\noindent in a~sequence, the identity
$\Sigma _{\theta ,\phi } \; \psi _{j_{\min }} \equiv  0$  holds.
The  case of negative $k$ can be considered in the same way.
Thus, at every $k$, the $j_{\min }$-state  equation  has  the  same
unique form
\begin{eqnarray}
\left (  i\; {\gamma ^{0} \over  \sqrt{\Phi}}  \; { \partial \over \partial t} \; + \; i\gamma ^{3} \sqrt{\Phi}\;
{\partial \over \partial r}\;   - \;  M
\right ) \psi _{j_{mi}} = 0 \; ;
\label{3.11c}
\end{eqnarray}

\noindent which leads to the same unique radial system

\vspace{3mm}

$
k = +1/2,+1,\ldots $
\begin{eqnarray}
{\epsilon \over \sqrt{\Phi}} \; f_{3} - i \; \sqrt{\Phi} { d\over dr}  \; f_{3}  - M \; f_{1} = 0\; ,
\nonumber
\\
{ \epsilon \over \sqrt{\Phi}} \; f_{1} + i \; \sqrt{\Phi} { d \over dr} \; f_{1}  - M \; f_{3} = 0 \; ;
\label{3.12a}
\end{eqnarray}

$
k = -1/2,-1,\ldots $
\begin{eqnarray}
{\epsilon  \over \sqrt{\Phi}} \; f_{4} + i \; \sqrt{\Phi} { d\over dr}\; f_{4} - M \;f_{2} = 0 \; ,
\nonumber
\\
{ \epsilon \over \sqrt{\Phi}}\; f_{2} - i \; \sqrt{\Phi} { d\over dr}\; f_{2} - M \;f_{4} = 0 \; .
\label{3.12b}
\end{eqnarray}

 In the limit of flat space--time, these equations are equivalent respectively to

$
k = + 1/2,+ 1,\ldots$
\begin{eqnarray}
\left ( {d^{2} \over dr^{2}}  + \epsilon ^{2}  - m^{2}\right ) f_{1}  = 0 \;
, \;\; f_{3} =  { 1 \over m}\left ( \epsilon  +
i { d \over dr }\right )  f_{1}     \; ;
\label{3.13a}
\end{eqnarray}
$ k = - 1/2, - 1,\ldots$
\begin{eqnarray}
\left ( {d^{2} \over dr^{2}}  + \epsilon ^{2}  - m^{2}\right )  f_{4} = 0\;
 ,\;\; f_{2} = {1 \over m} \left ( \epsilon  + i {d \over dr} \right )
 \; f_{4} \; .
\label{3.13b}
\end{eqnarray}

\noindent These equation  both lead us to the functions
 $f = \exp  ( \pm  \sqrt{m^{2} - \epsilon ^{2}} \; r )$.
In particular,  at $\epsilon \; < \; m$, we have a solution
\begin{eqnarray}
\exp  \; ( \; -\sqrt{m^{2} -\epsilon ^{2}} \; r \; ) \; ,
\label{4.13c}
\end{eqnarray}

\noindent which seems  to  be appropriate to describe   bound   states   in
the~electron-monopole system.

\section{  Solution of the radial equations}

 Let us turn back to the system
(\ref{3.7}) and (for definiteness)  consider equations  at  $\delta =
+1$ (formally the second case $\delta =-1$ corresponds to the
 change  $M \Longrightarrow - M$)
\begin{eqnarray}
( \sqrt{\Phi} {d \over dr} \;+\; {\nu \over r}\;) \; f \; + \; ( {
\epsilon  \over \sqrt{\Phi}}  \;+ \;  M )\; g \; = \;0 \; ,
 \nonumber
\\
( \sqrt{\Phi} {d \over dr} \; - \;{\nu \over r}\;)\; g  \;- \; (
{\epsilon  \over \sqrt{\Phi}}\; - \;
  M )\; f\; =\; 0     \; .
\label{10.12}
\end{eqnarray}

\noindent
 Here we see  additional singularities at the points
$$
 \epsilon +
\sqrt{\Phi} \;  M  =0   \qquad  \mbox{or} \qquad  \epsilon - \sqrt{\Phi} \;   M
= 0 \;.
$$
For instance, the equation for $f(r)$ has  the form
\begin{eqnarray}
{d^{2}  \over dr^{2}} f + \left ( {2r \over 1 +r^{2} } - {M r
\over  \sqrt{1+r^{2}} (\epsilon + M \sqrt{1 +r^{2}})} \right ) {d
\over dr } f + \left ( { \epsilon^{2} \over (1+r^{2})^{2} } -
{M^{2} \over 1 +r^{2}}  \right. \nonumber
\\
\left. -{ \nu^{2} \over r^{2} (1 + r^{2} ) } - {\nu \over
r^{2}(1+r^{2})^{3/2} } -
 { M \nu  \over  (1+r^{2})   (\epsilon + M \sqrt{1 +r^{2}}) }
\right ) f = 0 \; . \nonumber
\end{eqnarray}

However, there exists possibility to move these singularities away
through a special transformation of the functions $f(r), g(r)$
\cite{Otchik-1985}.
To this end,  let us
introduce a new variable $ r = \sinh\,\rho  $, eqs. (\ref{10.12})
look simpler
\begin{eqnarray}
( {d \over d \rho} + {\nu \over \sinh\, \rho})  f  + ( {
\epsilon  \over \cosh\, \rho }  +   M ) g  = 0 \; ,
 \nonumber
\\
(  {d \over d \rho}  - {\nu \over \sinh\, \rho}) g  -  (
{\epsilon  \over \cosh\, \rho } -
 M ) f = 0     \; .
\label{10.13}
\end{eqnarray}

\noindent Summing and subtracting two  last equations, we get
\begin{eqnarray}
{d \over d \rho} (f+g) + {\nu \over \sinh\,\rho} (f-g) -  {\epsilon
\over \cosh\, \rho} (f-g) + M (f+g) = 0 \; , \nonumber
\\
{d \over d \rho} (f-g) + {\nu \over \sinh\, \rho} (f+g) +  {\epsilon
\over \cosh\, \rho} (f+g) - M(f-g) = 0 \; . \label{10.14}
\end{eqnarray}

\noindent Introducing two  new  functions
\begin{eqnarray}
f + g = e^{-\rho/2} (F + G) \; , \qquad f - g = e^{+\rho/2} (F -
G) \; , \label{10.15}
\end{eqnarray}

\noindent or in matrix form
\begin{eqnarray}
\left | \begin{array}{c}
G \\ H
\end{array} \right | =
\left | \begin{array}{cc}
\cosh \rho /2  & - \sinh \rho /2 \\
- \sinh \rho /2 &  \cosh \rho /2 \\
\end{array} \right |
\left | \begin{array}{c}
g \\ h
\end{array} \right | ,
\label{Trans-a}
\end{eqnarray}

\noindent where (see definition of the variable $z$ below)
\begin{eqnarray}
\cosh {\rho \over 2} = \sqrt{{ \sqrt{1-z} +1\over 2}}\;  , \qquad
\sinh {\rho \over 2}  = \sqrt{{ \sqrt{1-z}-1 \over 2}} \; ,
\label{Trans-b}
\end{eqnarray}

\noindent one transforms (\ref{10.14})  into
\begin{eqnarray}
{d \over d \rho} e^{-\rho/2} (F + G)  + {\nu \over \sinh \rho}
e^{+\rho/2} (F - G) \nonumber
\\
  -{\epsilon \over \cosh \rho}  e^{+\rho /2} (F - G)  + M e^{-
\rho /2} (F + G)  = 0 \; , \nonumber
\\
{d \over d \rho} e^{+\rho /2} (F - G)  + {\nu \over \sinh \rho}
e^{-\rho /2} (F + G) \nonumber
\\
 + {\epsilon \over \cos \rho} e^{- \rho /2} (F + G)  - M e^{+
\rho /2} (F - G)  = 0 \; , \nonumber
\end{eqnarray}

\noindent or
\begin{eqnarray}
 {d \over d \rho}  (F + G)  - {1 \over 2} (F + G) +
{\nu \over \sinh \rho} (\cosh \rho +  \sinh \rho)  (F - G)
 \nonumber
\\
  -{\epsilon \over \cosh \rho}  (\cosh \rho +  \sinh \rho)  (F - G)
+ M  (F + G)  = 0 \; , \nonumber
\end{eqnarray}
\begin{eqnarray}
{d \over d \rho}  (F - G)  + {1 \over 2}  (F - G)  + {\nu \over
\sinh \rho}  (\cosh \rho - \sinh \rho) (F + G)
 \nonumber
\\
  +{\epsilon \over \cosh \rho} (\cosh \rho -  \sinh \rho) (F + G)  -
M  (F - G)  = 0 \; . \nonumber
\end{eqnarray}

\noindent Now summing and subtracting two last  equations, we obtain
\begin{eqnarray}
 ({d  \over d \rho}    + \nu \;  { \cosh \rho \over \sinh \rho}   -  \epsilon \; { \sinh \rho \over  \cosh  \rho}
 ) \; F
+ \;  ( \; \epsilon  + M  -  \nu   - {1 \over 2}   )\;  G = 0 \;
, \nonumber
\\
({d  \over d \rho}    - \nu   \; { \cosh \rho \over \sinh \rho}
 +  \epsilon \; { \sinh \rho \over  \cosh  \rho}  ) \; G
+ ( - \epsilon  + M   +  \nu   - {1 \over 2}  )  \; F = 0 \; .
\label{10.17}
\end{eqnarray}

Let us translate eqs. (\ref{10.17}) to the variable $z$:
\begin{eqnarray}
r^{2} = \mbox{sinh}^{2} \rho = - z, \qquad {d \over d \rho} =
2 \sqrt{-z(1-z)} {d \over dz} \; , \nonumber
\end{eqnarray}
\begin{eqnarray}
 \left ( 2\sqrt{-z(1-z)} {d \over dz}     + \nu   { \sqrt{1 - z}  \over \sqrt{-z} }   -
   \epsilon \; { \sqrt{-z}  \over  \sqrt{1 - z} }
 \right )  F
\nonumber
\\
+(  \epsilon  + M  -  \nu   - {1 \over 2}   )  G = 0 \; ,
\nonumber
\\
\left ( 2 \sqrt{-z(1-z)} {d \over dz}    - \nu    {
\sqrt{1 - z} \over \sqrt{-z} }
 +  \epsilon \; { \sqrt{-z} \over  \sqrt{1 - z} }  \right )  G
\nonumber
\\
+( - \epsilon  + M   +  \nu   - {1 \over 2}  )   F = 0 \; .
\label{10.17'}
\end{eqnarray}

\noindent
From  (\ref{10.17'}) it follow two 2-nd order differential equations
for  $F$ and  $G$ respectively
\begin{eqnarray}
z(1-z){d^{2}F\over dz^{2}}+ ({1\over 2}-z )
{dF\over dz}
\nonumber
\\
+\left[{1\over 4}\left(M-{1\over 2}\right)^{2}-{\epsilon(\epsilon-1)\over 4(1-z)}-
{\nu(\nu+1)\over 4z}\right]F=0\,,
\nonumber \\
z(1-z){d^{2}G\over dz^{2}}+ ({1\over 2}-z ){dG\over dz}
\nonumber
\\
+ \left[{1\over 4}\left(M-{1\over 2}\right)^{2}-
{\epsilon(\epsilon+1)\over 4(1-z)}-{\nu(\nu-1)\over 4z}\right]G=0\,.
\label{M10.17}
\end{eqnarray}

\noindent
With the use of substitutions
\begin{eqnarray}
F=z^{A}(1-z)^{B}\bar{F}(z)\,, \qquad
G=z^{K}(1-z)^{L}\bar{G}(z)\,,
\nonumber
\end{eqnarray}

\noindent eqs.  (\ref{M10.17}) take the form
\begin{eqnarray}
z(1-z)\,{d^{2} \bar{F}\over dz^{2}}+\left[2A+{1\over 2}-(2A+2B+1)z\right]\,{d\bar{F}\over dz}
\nonumber
\\
+ \left[{1\over 4}\left(M-{1\over 2}\right)^{2}-
(A+B)^{2}-{\epsilon(\epsilon-1)+2B(1-2B)\over 4(1-z)} \right.
\nonumber
\\
\left.
-  {\nu(\nu+1)-2A(2A-1)\over 4z}\right] \bar{F}=0\,,
\label{M1}
\end{eqnarray}

\vspace{3mm}
\begin{eqnarray}
z(1-z)\,{d^{2}\bar{G}\over dz^{2}}+\left[2K+{1\over 2}-(2K+2L+1)z\right]\,{d\bar{G}\over dz}
\nonumber
\\
+ \left[{1\over 4}\left(M-{1\over 2}\right)^{2}-
(K+L)^{2}-{\epsilon(\epsilon+1)+2L(1-2L)\over 4(1-z)} \right.
\nonumber
\\
\left.
- {\nu(\nu-1)-2K(2K-1)\over 4z}\right]\bar{G}=0\,.
\label{M2}
\end{eqnarray}

First let us consider eq.  (\ref{M1}); at $A$ and $B$ taken accordingly
\begin{eqnarray}
A={1+\nu\over 2}\,,\;-{\nu\over 2}\,,\qquad B={\epsilon\over 2}\,,\;{1-\epsilon\over 2}
\label{M3a}
\end{eqnarray}

\noindent it becomes simpler
\begin{eqnarray}
z(1-z)\,{d^{2}f\over dz^{2}}+\left[2A+{1\over 2}-(2A+2B+1)z\right]\,{df\over dz}
\nonumber
\\
+ \left[{1\over 4}\left(M-{1\over 2}\right)^{2}-(A+B)^{2}\right]f=0\,,
\label{M3b}
\end{eqnarray}

\noindent which is of hypergeometric type with parameters
$$
a={M\over 2}-{1\over 4}+A+B\,,\qquad b=-{M\over 2}+{1\over 4}+A+B\,, \qquad  c = 2A +1/2 \; .
$$

To construct functions appropriate to describe  bound states we must choose
\begin{eqnarray}
A={1+\nu\over 2}> 0 \,,\qquad B={1-\epsilon\over 2} < 0 \,, \qquad c = \nu +3/2 \; ;
\label{M3c}
\end{eqnarray}

\noindent  polynomial solutions  will arise with the  quantization rule imposed
\begin{eqnarray}
a=-n\,,\qquad \epsilon_{n} =M+2n +\nu +{3\over 2} \,,
\nonumber
\\
b = -n - M -1 / 2 \;, \qquad c =  \nu +3/2 \;.
\label{M3d}
\end{eqnarray}

Now let us turn  to eq. (\ref{M2}). At   $A,\;B$ chosen according to
\begin{eqnarray}
K={1-\nu\over 2}\,,\;{\nu\over 2}\,,\qquad L=-{\epsilon\over 2}\,,\;{1+\epsilon\over 2}
\label{M4a}
\end{eqnarray}

\noindent it will be simpler
\begin{eqnarray}
z(1-z)\,{d^{2}g\over dz^{2}}+\left[2K+{1\over 2}-(2K+2L+1)z\right]\,{dg\over dz}
\nonumber
\\
+ \left[{1\over 4}\left(M-{1\over 2}\right)^{2}-(K+L)^{2}\right]g=0\,,
\label{M4b}
\end{eqnarray}

\noindent which is of hypergeometric type
\begin{eqnarray}
\alpha={M\over 2}-{1\over 4}+K+L\,,\qquad \beta=-{M\over 2}+{1\over 4}+K+L\,, \qquad \gamma = 2K +{1\over 2} \; .
\nonumber
\end{eqnarray}

Again, to get bound states  we choose the values
\begin{eqnarray}
K={\nu\over 2}> 0 \,,\qquad L=-{\epsilon\over 2} < 0 \,,
\label{M4c}
\end{eqnarray}

\noindent then the quantization rule arises
\begin{eqnarray}
\alpha=-N \,,\qquad \epsilon_{N} =M   + 2 N +\nu -{1\over 2} \,.
\label{M4d}
\end{eqnarray}
It can be noted that
$
\epsilon_{N} = \epsilon_{n}$, when $ N = n +1$.

Let us calculate relative  coefficient between functions $F(z)$ and $G(z)$.
These being taken in the form
\begin{eqnarray}
F(z) =F_{0}\,z^{(1+\nu)/2 }\,(1-z)^{(1 -\epsilon)/ 2} \bar{F}(a,b,c;z)\,,\qquad c={3\over 2}+\nu \;,
\nonumber
\\
a={M\over 2}+{3\over 4}+{\nu\over 2}-{\epsilon\over 2}\,,
\qquad
 b=-{M\over 2}+{5\over 4}+{\nu\over 2}-{\epsilon\over 2}\,;
\label{M5}
\end{eqnarray}

\noindent and
\begin{eqnarray}
G(z) =G_{0}\,z^{ \nu / 2 }\,(1-z)^{-\epsilon / 2}\bar{G}(\alpha,\beta,\gamma;z)\,,\;\;
\gamma={1\over 2}+\nu= c- 1 \; ,
\nonumber
\\
\alpha={M\over 2}-{1\over 4}+{\nu\over 2}-{\epsilon\over 2}=
a-1 \,,\qquad
 \beta=-{M\over 2}+{1\over 4}+{\nu\over 2}-{\epsilon\over 2}=b-1 \,,
\label{M6}
\end{eqnarray}

\noindent
 must obey the following system
\begin{eqnarray}
 \left ( 2\sqrt{-z(1-z)} {d \over dz}     + \nu   { \sqrt{1 - z}  \over \sqrt{-z} }   -
   \epsilon \; { \sqrt{-z}  \over  \sqrt{1 - z} }
 \right )  F
+   (  + \epsilon  + M  -  \nu   - {1 \over 2}   )  G = 0 \; ,
\nonumber
\\
\left (2\sqrt{-z(1-z)} {d \over dz}     - \nu   \; {
\sqrt{1 - z} \over \sqrt{-z} }
 +  \epsilon  { \sqrt{-z} \over  \sqrt{1 - z} }  \right )  G
+ ( - \epsilon  + M   +  \nu   - {1 \over 2}  )   F = 0 \; .
\nonumber
\end{eqnarray}

\noindent To find a relative factor, it is convenient to  use the second equation
\begin{eqnarray}
\left (-2\sqrt{-z(1-z)} {d \over dz}    - \nu    {
\sqrt{1 - z} \over \sqrt{-z} }
 +  \epsilon  { \sqrt{-z} \over  \sqrt{1 - z} }  \right )  G
\nonumber
\\
 + ( - \epsilon  + M   +  \nu   - {1 \over 2}  )   F = 0 \,.
\nonumber
\end{eqnarray}

\noindent
Substituting expressions for  $F$ and  $G$, after simple calculation we get to
\begin{eqnarray}
2\, i G_{0} \,{d\bar{G}\over dz}=F_{0} \,(-\epsilon+M+\nu- {1\over 2})\,\bar{F} \; .
\nonumber
\end{eqnarray}

\noindent
Allowing for the known rule for differentiating hypergeometric functions
\begin{eqnarray}
{ d \over  d z }  \bar{G} (z) = { d  \over  dz }  F( a-1, b-1, c-1; z)
\nonumber
\\
= {(a-1)(b-1) \over c-1} F(a,b,c ; z) = {(a-1)(b-1) \over c-1}  \bar{F}(z) \; ,
\nonumber
\end{eqnarray}

\noindent we  obtain
$$
2\, i G_{0} \, {(a-1)(b-1) \over c-1} =F_{0} \,(-\epsilon+M+\nu- {1\over 2}) \; .
$$

\noindent
Ultimately,
we arrive at the formula
\begin{eqnarray}
F_{0} =  i  \,  { M-1/2 + N \over 2}   \; G_{0} \; ,
\label{A}
\end{eqnarray}

\noindent remembering that
$
\epsilon_{N} =M - 1/2  + 2 N +\nu $.

\section{  Radial equations in the case $j_{min}$ }

Let us turn back to the case of the minimal value of $j$:

\vspace{3mm}

$
k = +1/2,+1,\ldots $
\begin{eqnarray}
{\epsilon \over \sqrt{\Phi}} \; f_{3} - i \; \sqrt{\Phi} { d\over dr}  \; f_{3}  - M \; f_{1} = 0\; ,
\nonumber
\\
{ \epsilon \over \sqrt{\Phi}} \; f_{1} + i \; \sqrt{\Phi} { d \over dr} \; f_{1}  - M \; f_{3} = 0 \; ;
\label{5.1a}
\end{eqnarray}

\noindent from where for new functions
\begin{eqnarray}
H \; = \; {f_{1} + f_{3} \over \sqrt{2}} \; , \qquad G \; = \;
{f_{1} - f_{3} \over i \sqrt{2}} \; \nonumber
\end{eqnarray}

\noindent
we derive

\vspace{3mm}
$
k = +1/2,+1,\ldots $
\begin{eqnarray}
\sqrt{\Phi} {d \over dr} H + ( {\epsilon \over \sqrt{\Phi}} + M) G = 0\; ,
\nonumber
\\
\sqrt{\Phi} {d \over dr} G - ( {\epsilon \over \sqrt{\Phi}} - M) H = 0 \; .
\label{5.1b}
\end{eqnarray}

And in the same manner for another case we have

\vspace{3mm}
$
k = -1/2,-1,\ldots $
\begin{eqnarray}
{\epsilon  \over \sqrt{\Phi}} \; f_{4} + i \; \sqrt{\Phi} { d\over dr}\; f_{4} - M \;f_{2} = 0 \; ,
\nonumber
\\
{ \epsilon \over \sqrt{\Phi}}\; f_{2} - i \; \sqrt{\Phi} { d\over dr}\; f_{2} - M \;f_{4} = 0 \; ,
\label{5.2a}
\end{eqnarray}

\noindent from whence  for new functions
\begin{eqnarray}
H \; = \; {f_{2} + f_{4} \over \sqrt{2}} \; , \qquad G \; = \;
{f_{2} - f_{4} \over i \sqrt{2}} \; \nonumber
\end{eqnarray}
we obtain
\begin{eqnarray}
\sqrt{\Phi} {d \over dr} G + ( {\epsilon \over \sqrt{\Phi}} - M) H = 0\; ,
\nonumber
\\
\sqrt{\Phi} {d \over dr} H - ( {\epsilon \over \sqrt{\Phi}} + M) G = 0 \; .
\label{5.2b}
\end{eqnarray}

We can use  the above method to eliminate nonphysical singular
points.
Let us  perform special transformation on the  functions
\begin{eqnarray}
G + H = e^{-\rho /2 } (g + h) \;, \qquad
G - H = e^{+\rho /2 } (g - h) \; .
\label{5.4a}
\end{eqnarray}

\noindent
After simple calculation we arrive  at

\vspace{3mm}
\underline{instead of (\ref{5.1b})}

\begin{eqnarray}
 ({d  \over d \rho}     +  \epsilon  { \sinh \rho \over  \cosh  \rho}
 ) \; g
+   ( \; -\epsilon  + M     - 1 / 2   )\;  h = 0 \; , \nonumber
\\
({d  \over d \rho}
 -  \epsilon  { \sinh \rho \over  \cosh  \rho}  ) \; h
+ ( + \epsilon  + M     - 1 / 2  )  \; g = 0 \; ; \label{5.5}
\end{eqnarray}

\vspace{3mm}
\underline{instead of (\ref{5.2b})}

\begin{eqnarray}
({d \over d \rho}
 +  \epsilon \; { \sinh \rho \over  \cosh  \rho}  ) \; h
+ ( - \epsilon  - M     - 1 /  2  )  \; g = 0 \;,
 \nonumber
\\
 ({d  \over d \rho}     -  \epsilon  { \sinh \rho \over  \cosh  \rho}
 ) \; g
+ \;  ( + \epsilon  - M     - 1 /  2   )\;  h = 0 \;  .
\label{5.5'}
\end{eqnarray}

In the variable  $z$
\begin{eqnarray}
r= \sinh \rho = \sqrt{-z} \; \nonumber
\end{eqnarray}

\noindent the system  (\ref{5.5}) takes the form
\begin{eqnarray}
\sqrt{-z(1-z)}  \left ( {d \over dz} - {\epsilon /2 \over 1-z}
\right ) g - {(-\epsilon  + M  -1/2 )\over 2} \;h = 0 \; ,
\nonumber
\\
\sqrt{-z(1-z)}  \left ( {d \over dz} +{\epsilon /2 \over 1-z}
\right ) h - {(+ \epsilon  +M  -1/2 ) \over 2} \; g = 0 \; .
\label{5.6}
\end{eqnarray}

\noindent Note that the system is symmetric with respect to changes
\begin{eqnarray}
f \Longleftrightarrow  h \;, \qquad  \epsilon \Longleftrightarrow  - \epsilon \; .
\label{symmetric}
\end{eqnarray}

After excluding the function   $h$  from (\ref{5.6}) we get
\begin{eqnarray}
h = {2 \over (- \epsilon + M  -1/2) } \; \sqrt{(-z)(1-z)}  \left (
{d \over dz} - {\epsilon /2 \over 1-z} \right ) g \;  , \nonumber
\\[4mm]
\sqrt{(-z)(1-z)}  \left ( {d \over dz} +{\epsilon /2 \over 1-z}
\right ) \sqrt{(-z)(1-z)} \left ( {d \over dz} - {\epsilon /2
\over 1-z} \right ) g \nonumber \\
-  {(M-1/2)^{2} -  \epsilon^{2}) \over 4} \; g = 0 \; .
\label{5.7}
\end{eqnarray}

\noindent Ultimately, an equation  for $g(z)$ reads
\begin{eqnarray}
z(1-z)  {d^{2} g \over dz^{2} }  +   (1/2 - z  )   {d g\over dz} +
\left ( {(M- 1/2 )^{2} \over 4 }-
    {\epsilon^{2}+ \epsilon \over 4} \;  {1 \over 1-z}
   \right ) g = 0 \; .
\label{5.8}
\end{eqnarray}

In the same manner we get a second order differential equation for  $h$ after exclusion of $g$:
\begin{eqnarray}
g = {2 \over (+ \epsilon + M  -1/2) } \; \sqrt{(-z)(1-z)}  \left (
{d \over dz} + {\epsilon /2 \over 1-z} \right ) h \;  , \nonumber
\\[4mm]
\sqrt{(-z)(1-z)}  \left ( {d \over dz} -{\epsilon /2 \over 1-z}
\right ) \sqrt{(-z)(1-z)} \left ( {d \over dz} + {\epsilon /2
\over 1-z} \right ) h
 \nonumber \\
-  {(M-1/2)^{2} -  \epsilon^{2}) \over 4} \; h = 0 \; ,
\label{5.9}
\end{eqnarray}
and  ultimately
\begin{eqnarray}
z(1-z)  {d^{2} h \over dz^{2} }  +   (1/2 - z  )   {d h\over dz} +
\left ( {(M- 1/2 )^{2} \over 4 }-
    {\epsilon^{2}- \epsilon \over 4} \;  {1 \over 1-z}
   \right ) h = 0 \; .
\label{5.10}
\end{eqnarray}

Equations (\ref{5.10}) and (\ref{5.8}) differ only in the sign at the  parameter $\epsilon$.

\section{ Solutions of radial equations in the case $j_{min}$ }

With the use of  substitution
 $
g = (1-z)^{A} \varphi (z)
$,  from (\ref{5.8})   we produce  for $\varphi$
\begin{eqnarray}
z(1-z) \;\varphi'' + [ {1\over 2} - (1 + 2A) z ] \; \varphi'
\nonumber
\\
+\left  [ \left ( A^{2} -{A \over 2} - {\epsilon^{2} +\epsilon \over 4}
\right ) {1 \over 1- z} -A^{2}
 + {(M-1/2)^{2} \over 4}   \right ] \varphi\; .
\label{6.2}
\end{eqnarray}

\noindent
Requiring
\begin{eqnarray}
 A^{2} -{A \over 2} - {\epsilon^{2} +\epsilon \over 4}
 = 0\qquad \Longrightarrow \qquad 2A = \epsilon +1  , - \epsilon
\nonumber
\end{eqnarray}

\noindent
one gets
\begin{eqnarray}
z(1-z) \;\varphi'' + [ {1\over 2} - (1 + 2A) z ] \; \varphi'  - {4   A^{2}
 - (M-1/2)^{2} \over 4}   \;  \varphi =0 \; ,
\nonumber
\\
\varphi = F(a, b, c, z) \;, \; c = {1\over 2} \; ,
\;
 a + b = 2A , \; a b = {4   A^{2}
 - (M-1/2)^{2} \over 4} \; ,
\label{6.3}
\end{eqnarray}

\noindent
that is
\begin{eqnarray}
a = {2A + (M -1/2) \over 2}  \; , \qquad  b  = { 2A - (M -1/2) \over 2} \; .
\label{6.4}
\end{eqnarray}

\noindent
Below we will use negative values for $A$
\begin{eqnarray}
A = -  \epsilon  /2 \;, \qquad
g  (z) = (1-z)^{-\epsilon/2} \varphi (z) \; ;
\label{6.5}
\end{eqnarray}

\noindent
so that
\begin{eqnarray}
 a = {-\epsilon   + (M -1/2) \over 2}  \; , \qquad  b  = { -\epsilon   - (M -1/2) \over 2}\; .
\label{6.6}
\end{eqnarray}

Any 2-nd order differential equation has two linearly independent solutions; here they are
\begin{eqnarray}
\varphi _{1} =   U_{1}(z) = F(a,b,c; z)\; ,
\nonumber
\end{eqnarray}
\begin{eqnarray}
\varphi_{2} =  U_{5} (z) = z^{1-c} F(a+1-c,b+1-c,2-c; z)  \; .
\label{solution-1}
\end{eqnarray}

Similar analysis can be performed for eq. (\ref{5.10})
\begin{eqnarray}
z(1-z)  {d^{2} h \over dz^{2} }  +   (1/2 - z  )   {d h\over dz} +
\left ( {(M- 1/2 )^{2} \over 4 }-
    {\epsilon^{2}- \epsilon \over 4} \;  {1 \over 1-z}
   \right ) h = 0 \; .
\nonumber
\\
\label{6.12}
\end{eqnarray}

\noindent
With the use of  substitution
 $
h(z)  = (1-z)^{L} \eta (z)
$,  for $\eta(z)$  we produce
\begin{eqnarray}
z(1-z) \;\eta'' + [ {1\over 2} - (1 + 2L) z ] \; \eta'
\nonumber
\\
+\left  [ \left ( L^{2} -{L \over 2} - {\epsilon^{2} -\epsilon \over 4}
\right ) {1 \over 1- z} -L^{2}
 + {(M-1/2)^{2} \over 4}   \right ] \eta\; .
\label{6.13}
\end{eqnarray}

\noindent
Requiring
\begin{eqnarray}
 L^{2} -{L \over 2} - {\epsilon^{2} -\epsilon \over 4}
 = 0\qquad \Longrightarrow \qquad 2L = +\epsilon ,  - \epsilon + 1
\nonumber
\end{eqnarray}

\noindent
one gets
\begin{eqnarray}
z(1-z) \;\eta'' + [ {1\over 2} - (1 + 2L) z ] \; \eta'  - {4   L^{2}
 - (M-1/2)^{2} \over 4}   \;  \eta =0 \; ,
\nonumber
\\
\eta = F(\alpha, \beta, \gamma, z) \;, \qquad \gamma  = {1\over 2} \; ,
\nonumber
\\
 \alpha + \beta = 2L, \qquad \alpha \beta = {4   L^{2}
 - (M-1/2)^{2} \over 4} \; ,
\label{6.14}
\end{eqnarray}

\noindent
that is
\begin{eqnarray}
\alpha = {2L - (M -1/2) \over 2}  \; , \qquad  \beta  = { 2L + (M -1/2) \over 2} \; .
\label{6.15}
\end{eqnarray}

\noindent
Below we will use negative values for $L$
\begin{eqnarray}
L = (-  \epsilon +1)  /2 < 0  \;, \qquad
h  (z) = (1-z)^{(-\epsilon+1)/2} \eta (z)\; ,
\label{6.16}
\end{eqnarray}

\noindent
so that
\begin{eqnarray}
 \alpha  = {-\epsilon +1  + (M -1/2) \over 2}  \; , \qquad  \beta  = {-\epsilon +1  - (M -1/2) \over 2}  \;  .
\label{6.17}
\end{eqnarray}

Functions $g(z)$ and $h(z)$
must obey the above system of  first order differential equations.
To verify that,   let us start with the functions
\begin{eqnarray}
g = G_{0} (1-z)^{A} \varphi_{1}(z)\; , \qquad 2A = - \epsilon \; ,
\nonumber
\\
 \varphi_{1} =F(a,b,c,z) \qquad c= 1/2 \; ,
\nonumber
\\
a = {-\epsilon   + (M-1/2) \over 2} \; ,\qquad
 b = {-\epsilon   - (M-1/2) \over 2}\; ,
\end{eqnarray}
\begin{eqnarray}
h = H_{0}(1-z)^{L} \eta_{2}(z) \; , \qquad 2L = - \epsilon +1 \; , \qquad
\nonumber
\\
\eta_{2} = z^{1/2} F(\alpha+1 -\gamma \; , \beta +1 - \gamma, 2 - \gamma,z) \; ,
\nonumber
\\
\alpha +1 - \gamma  = {-\epsilon +2   + (M-1/2) \over 2}  = a +1\;    ,
\nonumber
\\
\beta +1 - \gamma  = {-\epsilon  +2  - (M-1/2) \over 2}  = b +1\; ,
\nonumber
\\
2- \gamma = c +1 \; ,
\end{eqnarray}

\noindent
and relate them  with the help of the first equation in the system (\ref{5.6})
\begin{eqnarray}
\sqrt{(-z)(1-z)}  \left ( {d \over dz} - {\epsilon /2 \over 1-z}
\right ) g - {(-\epsilon  + M  -1/2 )\over 2} \;h = 0  \; .
\nonumber
\end{eqnarray}

\noindent
After simple calculations we obtain
\begin{eqnarray}
G_{0} i  {d \over dz} F(a,b,c,z) = H_{0}  {(-\epsilon  + M  -1/2 )\over 2}
F( a+1,b+1,c+1,z) \; ,
\nonumber
\end{eqnarray}

\noindent
from whence it follows
\begin{eqnarray}
G_{0} i  {ab \over c}  = H_{0} {(-\epsilon  + M  -1/2 )\over 2} \; ,
\nonumber
\end{eqnarray}

\noindent  that is
$$
H_{0}=i\,(-\epsilon  - M  +1/2 )\,G_{0}\,.
$$

To get polynomial  solutions
we must require
\begin{eqnarray}
a = - n \; \;\; \Longrightarrow \;  \;\;  \epsilon_{n}  =  M + 2n  -1/2\;,
\nonumber
\\
b = -n - M + 1/  2  \; , \qquad c = 1/2 \; ,
\nonumber
\\
g  (z) = (1-z)^{-(\epsilon_{n}+1)/2} F(a, b, c , z) \; .
\label{spectrum}
\end{eqnarray}

\noindent note that
\begin{eqnarray}
g (z) = (1-z)^{-n - (M-1/2)/2} \;' F(-n, -n -M + {1 \over 2}, {1 \over   2} , z) \; ;
\end{eqnarray}

\noindent
therefore as $z = -r^{2} \longrightarrow - \infty$ the function $g(z)$ tends to zero
\begin{eqnarray}
g(z) \longrightarrow 0\;  , \qquad \mbox{only \; if } \qquad M > {1 \over 2} \; .
\nonumber
\end{eqnarray}

In usual units, that condition for existence of bound states consistent with anti de Sitter geometry structure,
the inequality   $M > {1 \over 2} $,  looks as
$$
\rho >{1\over 2} {\hbar \over Mc } =  {1\over 2} \lambda_{e}=  1.213 \times 10^{-12} \; metre
$$

\noindent so it can be  broken  only in a very strong anti de Sitter gravitation background, the latter is beyond of our treatment.

Let us write down several energy levels (in usual units)
\begin{eqnarray}
\epsilon_{0} = Mc^{2} -{1 \over 2} {c\hbar \over \rho} \; ,\qquad
\epsilon_{1} = Mc^{2} +{3 \over 2} {c\hbar \over \rho} \; ,\qquad
\epsilon_{2} = Mc^{2} +{5 \over 2} {c\hbar \over \rho} \; , \; ...
\nonumber
\\
\end{eqnarray}

or
\begin{eqnarray}
\epsilon_{0} = Mc^{2}(1  -{1 \over 2} {\lambda_{e}  \over \rho}) \; ,\qquad
\epsilon_{1} = Mc^{2} (1 +{3 \over 2} {\lambda_{e} \over \rho} )\; ,\qquad
\epsilon_{2} = Mc^{2}(1  +{5 \over 2} {\lambda_{e}  \over \rho})  \; , \; ...
\nonumber
\\
\end{eqnarray}

If one mentally increases  the curvature radius $\rho$, the energy levels will become   denser and the  minimal  level
tends to the value $Mc^{2}$
\begin{eqnarray}
\epsilon_{0} = Mc^{2}(1  -{1 \over 2} {\lambda_{e}  \over \rho}) \;\;  \longrightarrow \;\; Mc^{2} \; .
\end{eqnarray}

\section{  Conclusions and discussion}

To understand  better results, let us discuss the case
of minimal $j_{min}$  in
 the limit of vanishing curvature. To this end, let us specify in more detail solutions
 for minimal values $j_{min}$
in  Minkowski space:

\vspace{3mm}
$
k = +1/2,+1,\ldots $
\begin{eqnarray}
\epsilon  \; f_{3} - i \;  { d\over dr}  \; f_{3}  - M \; f_{1} = 0\; ,
\nonumber
\\
 \epsilon  \; f_{1} + i \;  { d \over dr} \; f_{1}  - M \; f_{3} = 0 \; ;
\label{C.5}
\end{eqnarray}

$
k = -1/2,-1,\ldots $
\begin{eqnarray}
\epsilon   \; f_{4} + i \;  { d\over dr}\; f_{4} - M \;f_{2} = 0 \; ,
\nonumber
\\
 \epsilon \; f_{2} - i \;  { d\over dr}\; f_{2} - M \;f_{4} = 0 \; .
\label{C.6}
\end{eqnarray}

Let detail the case of positive $
k = +1/2,+1,\ldots $. Let it be
\begin{eqnarray}
{f_{1} + f_{3} \over \sqrt{2} } = h(r) \; , \qquad  { f_{1} - f_{3} \over  i \sqrt{2} } = g(r)
\label{C.7}
\end{eqnarray}

\noindent
relevant equations are
\begin{eqnarray}
{d \over dr }h +( \epsilon + M) \; g = 0 \; , \qquad
{d \over dr} g - (\epsilon - M) h = 0 \; .
\label{C.8}
\end{eqnarray}

\noindent
With the substitutions
\begin{eqnarray}
h(r) = H e^{\gamma r}\;, \qquad g (r) = G e^{\gamma r}
\label{C.9}
\end{eqnarray}

\noindent
we get (first let it be $(\epsilon^{2} - M^{2}) > 0 $)
\begin{eqnarray}
\gamma^{2} = - (\epsilon^{2} - M^{2}) \equiv = -  p ^{2}  \;, \qquad \gamma = +i p, - i p \; .
\nonumber
\\
H \gamma + (\epsilon + M) G =0\; \qquad \mbox{or} \qquad
G \gamma - (\epsilon - M) H = 0 \; .
\label{C.10}
\end{eqnarray}

\noindent
Thus we have two linearly independent solutions
\begin{eqnarray}
h_{1}(r) = H_{1} e^{+ip r} \; , \qquad g_{1}(r) = G_{1} e^{+ipr} \; ,
\qquad G_{1} = {\epsilon - M \over  ip} H_{1} \; ;
\nonumber
\\
h_{2}(r) = H_{2} e^{-ip r} \; , \qquad g_{2}(r) = G_{2} e^{-ipr} \; , \qquad G_{2} = {\epsilon - M \over
 - ip} H_{2} \; .
\label{C.11}
\end{eqnarray}

Below, we take $H_{1}=H_{2} = 1$. We can introduce two linear combinations of these solutions

the first
\begin{eqnarray}
{ h_{1}(r) + h_{2}(r) \over 2 }= \cos pr \;,
\nonumber
\\
{ g_{1}(r) + g_{2}(r) \over 2 }=
{\epsilon - M \over  p} \;  \sin pr \; ;
\label{C.12}
\end{eqnarray}

the second
\begin{eqnarray}
{ h_{1}(r) - h_{2}(r) \over 2 i }= \sin pr \;,
\nonumber
\\
{ g_{1}(r) - g_{2}(r) \over 2 i }=
{\epsilon - M \over - p} \;  \cos pr \; .
\label{C.13}
\end{eqnarray}

Now let us specify the case
 $(\epsilon^{2} - M^{2}) < 0 $)
\begin{eqnarray}
\gamma^{2} = - (\epsilon^{2} - M^{2}) \equiv = +  q ^{2}  \;, \qquad \gamma = + q, - q \; .
\nonumber
\\
H \gamma + (\epsilon + M) G =0\; \qquad \mbox{or} \qquad
G \gamma - (\epsilon - M) H = 0 \; .
\label{C.14}
\end{eqnarray}

Thus we have two linearly independent solutions
\begin{eqnarray}
h_{1}(r) = H_{1} e^{+q r} \; , \qquad g_{1}(r) = G_{1} e^{+qr} \; ,
\qquad G_{1} = {\epsilon - M \over  q} H_{1} \; ;
\nonumber
\\
h_{2}(r) = H_{2} e^{-q r} \; , \qquad g_{2}(r) = G_{2} e^{-q r} \; , \qquad G_{2} = {\epsilon - M \over  - q} H_{2} \; .
\label{C.15}
\end{eqnarray}

Below, we take $H_{1}=H_{2} = 1$. We can introduce two linear combinations of these solutions

the first
\begin{eqnarray}
{ h_{1}(r) + h_{2}(r) \over 2 }= \mbox{cosh}\;  qr \;,
\nonumber
\\
{ g_{1}(r) + g_{2}(r) \over 2 }=
{\epsilon - M \over  q} \;  \mbox{sinh}\;  qr
\label{C.16}
\end{eqnarray}

the second
\begin{eqnarray}
{ h_{1}(r) - h_{2}(r) \over 2  }=  \mbox{sinh}\;  qr \;,
\nonumber
\\
{ g_{1}(r) - g_{2}(r) \over 2  }=
{\epsilon - M \over  q} \;  \mbox{cosh}\;  qr \; .
\label{C.17}
\end{eqnarray}

Evidently, above  constructed  solutions in de Sitter model provide us with  generalizations of
these of  Minkowski space.
 It may be verified additionally by
direct limiting process when $\rho \rightarrow \infty$.
To this end, let us translate  solutions in de Sitter space to usual units
\begin{eqnarray}
g_{1}(R)  =   \left ( 1 +{R^{2}\over \rho^{2} } \right )^{-{E\rho \over  2c \hbar }}
 F(a,b,c;  -{R^{2} \over \rho^{2}}) \;, \qquad c = 1/2\; ,
\nonumber
\\
g_{2}(R)  =  R    \left ( 1 +{R^{2}\over \rho^{2} } \right )
^{-{E\rho \over 2 c \hbar} }   F(a+1-c,b+1-c,2-c; -{R^{2} \over \rho^{2}})  \; ,
\nonumber
\\
h_{1} (R)  =  \left ( 1 +{R^{2}\over \rho^{2} } \right )^{-{E\rho \over 2c \hbar} +1/2 }
F(\alpha ,\beta, \gamma ; -{R^{2} \over \rho^{2}})  \; , \qquad \gamma = 1/2 \; ,
\nonumber
\\
h_{2}(R)  = R     \left ( 1 +{R^{2}\over \rho^{2} } \right )^{-{E\rho \over  2c \hbar} +1/2 }
F(\alpha +1-\gamma,\beta +1-\gamma ,2-\gamma; -{R^{2} \over \rho^{2}})\; ,
\nonumber
\end{eqnarray}

\noindent
Parameters of hypergeometric functions are given by
\begin{eqnarray}
a = {1 \over 2} \left  (  - {E \rho \over c \hbar }   + ({mc\rho \over \hbar} - {1 \over 2})  \right   )  ,
\qquad
b =   {1 \over 2} \left (  - {E \rho \over c \hbar } - ({mc\rho \over \hbar} - {1 \over 2} )    \right )  ,
\nonumber
\end{eqnarray}

\begin{eqnarray}
\alpha  = {1 \over 2} \left  (  - {E \rho \over c \hbar }  +1  + ({mc\rho \over \hbar} - {1 \over 2})  \right   )   ,
\qquad
\beta =    {1 \over 2} \left  (  - {E \rho \over c \hbar }  +1 - ({mc\rho \over \hbar} - {1 \over 2})  \right   )  .
\nonumber
\end{eqnarray}

Let us examine the limiting procedure  at $\rho \rightarrow \infty$ in
$
 F(a,b,c;  -R^{2} /  \rho^{2}) \; .
$
Because
\begin{eqnarray}
{1 \over 1!} \; {ab \over c}  \; (- {R^{2} \over \rho^{2}}) \;\;  \rightarrow  \;\;  {1 \over 2!}  ( m^{2}c^{2} /
 \hbar^{2} -E^{2} / \hbar^{2} c^{2} )R^{2}  = - {1 \over 2!} \; (pR)^{2}\; ,
\nonumber
\\
{1 \over 2!} \; {a(a+1)b (b+1)  \over c(c+1)} \; (-   {R^{2} \over \rho^{2}} )\;\;  \rightarrow
 \;\;    + { (pR)^{4} \over  4! } \; ,
\nonumber
\\
{1 \over 3!} \; {a(a+1)(a+2)b (b+1) (b+2) \over c(c+1)(c+2)} \; (- {R^{2} \over \rho^{2}}) \;\;  \rightarrow
 \;\;    - { (pR)^{6} \over  6! } \; ,
\nonumber
\end{eqnarray}

\noindent and so on, we obtain  the following limiting relation
\begin{eqnarray}
\lim_{\rho \infty }  F(a,b,c;  - {R^{2} \over \rho^{2}})  = \cos pr \qquad \Longrightarrow
\qquad \lim_{\rho \infty }  g_{1} (R)   = \cos pr \;.
\nonumber
\end{eqnarray}

Similarly, we get
\begin{eqnarray}
\lim_{\rho \infty }  h_{1} (R)   = \cos pr \; .
\end{eqnarray}

In the same manner,   we arrive at two limiting  relationships
\begin{eqnarray}
\lim _{\rho \infty } \; p R      \; g_{2} (R) = \sin p R \;, \qquad
\lim _{\rho \infty } \; p R      \; h_{2} (R)  = \sin p R \;.
\end{eqnarray}

To rationalize how the finite sums  (polynomials of $n$-order )
may approximate the  functions $\cos pR $ and $\sin  pR$ (infinite series), we should take into account
the quantization condition
$$
 \alpha = - n \qquad \Longrightarrow \qquad
  E = Mc^{2} + (2n -{1\over 2}) {c \hbar \over  \rho}
  $$

\noindent
At any fixed $E$, as $\rho$ increases the number $n$ also must increase. This means,
 that the finite  sums of $n$ terms when $\rho  $ increases
will approximate infinite series.

\section{Acknowledgements}

Authors are  grateful  to  V.M. Red'kov  for  moral support and  advices.
This  work was   supported   by the Fund for Basic Researches of Belarus
 F11M-152.

\end{document}